# Magnetic Property of Rutile $Ti_{0.94}A_{0.06}O_2$ (A=Li, Mg, K) Compounds

S. K. Srivastava[1*], M. Brahma[1], B. Brahma[1] and G. Bouzerar[2]

[1]*Department of Physics, Central Institute of Technology, Kokrajhar, Assam, India*
[2]*Institut Néel, CNRS Grenoble and Université Joseph Fourier, Grenoble, France*
*E-mail: sk.srivastava@cit.ac.in*

**Abstract.** Our purpose is to study role of ionic radii of non-magnetic dopants; $Li^+$ (0.68 Å), $Mg^{2+}$ (0.72 Å) and $K^+$ (1.38 Å) on the magnetic property of rutile $TiO_2$ compound. The $Ti_{0.94}A_{0.06}O_2$ (A=Li, Mg, K) compounds have been synthesized via solid state route method *at equilibrium*. The structural analyses of X-ray diffraction pattern reveals that the doping of Li and Mg lead to Ti site substitution and K doping lead to core shell kind of structure. The magnetic property measurement by SQUID magnetometer indicate that all compounds exhibit weak paramagnetism with highest paramagnetic moment of ~ 0.3 $\mu_B$ / ion for K doped compound but no long-range ferromagnetic ordering. We have discussed the observed magnetism in correlation with the nature of substitution.

## INTRODUCTION

In recent years, one new kind of materials so called $d^0$ ferromagnetic oxides ($A_{1-x}B_xO_2$) have been largely emphasized as the alternative materials to transition metal (TM) doped ferromagnetic semi-conductors. The interest in such materials consists in the use of new devices where spin degrees of freedom carry information in order to reduce electrical consumption, allow non-volatility, store and manipulate data, beyond room temperature (RT). The $d^0$ or intrinsic ferromagnetism had been reported in several oxides such as $HfO_2$, CaO, ZnO, $CaB_6$ and $ZrO_2$ [1-5]. From a theoretical point of view, it has been shown that point defects as cation vacancies may be at the origin of the magnetism of these materials [2, 6-8]. In a recent model, which includes disorder and electron-electron correlation effects on equal footing, it has been proposed that high Curie temperatures could be reached in dioxides such as $AO_2$ ($A$=Ti, Zr, or Hf) by substitution of a monovalent cation of the group 1A [9]. Following this model, recent ab initio studies have predicted ferromagnetism with $T_C$ far beyond room temperature in these dioxides such as K-$SnO_2$ [10], Mg–$SnO_2$ [11], anatase Li-$TiO_2$ [12], rutile K–$TiO_2$ [13], V-$TiO_2$ [14], and K–$ZrO_2$ [5, 13]. Experimentally, RT ferromagnetism has been claimed in Cu doped $TiO_2$ prepared in thin film [15, 16] and carbon-doped $TiO_2$ prepared by solid state route [17] and Li doped $SnO_2$ [18]. Recent experiments by our group [19-20] extended the possibility of $d^0$ magnetism to cases where non-magnetic ions are not located at the substitution sites assumed in ab-initio studies.

$TiO_2$ is a transparent wide band gap material, with applications in photocatalysis, solar cells, semiconducting gas sensors, and as building blocks for photonic crystals. Our purpose is to study role of ionic radii of non-magnetic dopants; $Li^+$ (0.68 Å), $Mg^{2+}$ (0.72 Å) and $K^+$ (1.38 Å) on the magnetic property of rutile $TiO_2$ compound. The substitution of $Ti^{4+}$ by 6% of dopant provides enough numbers of holes to induce $d^0$ magnetism, as it has been already demonstrated experimentally [20]. We have prepared $Ti_{0.94}A_{0.06}O_2$ (A= Li, Mg, K) compounds to study the $d^0$ magnetism in correlation with the nature of substitution.

## EXPERIMENTAL TECHNIQUES

We have prepared $Ti_{0.94}A_{0.06}O_2$ (A= Li, Mg, K) compounds by standard solid state route method using high-purity $TiO_2$ (purity, 99.995 %), $Li_2CO_3$ (99.998 %), $Na_2CO_3$ (99.99 %) and $K_2CO_3$ (99.99 %) compounds. The maximum amount of any kind of trace magnetic impurities in the starting materials was found to be less than 0.9 ppm (parts per million) as mentioned by the supplier chemical analyses report. The final annealing in pellet form was carried out at 1200 $^{0}$C for 30 hrs in air. Slow scan powder X-Ray diffraction patterns were collected by using Philips XRD machine with CuK$_{\alpha}$ radiation. Magnetization (M) measurements as a function of magnetic field (H) and temperature (T) were carried out using a commercial SQUID magnetometer (Quantum Design, MPMS).

## RESULTS AND DISCUSSIONS

The X-Ray diffraction patterns for all samples are shown in Figure 1. All the diffractions peaks could be indexed on the basis of the tetragonal rutile type-structure. No extra diffraction peaks were detected showing that no crystalline parasitic phases are present in the samples within the limit of XRD. One typical refinement of XRD pattern with Fullprof program of $Ti_{0.94}Mg_{0.06}O_2$ compound is shown in Figure 2. The refinement of X-Ray patterns reveals no significant change in the lattice parameters for Li doped compound, in comparison to pure $TiO_2$ (a = b = 4.5945 Å and c = 2.9586 Å). However Mg doped compound show a small increase i.e. a = b = 4.604 Å and c = 2.958 Å, which seems to be in agreement, as ionic radii of $Mg^{2+}$ is little larger than $Ti^{4+}$. On contrary, no significant change in the lattice parameters was found for K doped compound, indicating that K ions do not substitute Ti ions. This can be understood by the large difference between the ionic radii of the 6-coordinate $Ti^{4+}$ (0.61 Å) and that of $K^{+}$ (1.38 Å). The oxidization state of Ti ion is found to be in $Ti^{4+}$, as it was confirmed from the XPS measurement [20].

The M-H measurement with SQUID magnetometer for $TiO_2$ compound shows that it exhibit weak paramagnetic behavior with a value of 0.005 emu/gm at 3K and at 5 Tesla field. However, the magnetization for the doped sample (Figure 3a) increases almost 10 times and it is maximum for the case of K doped compound. Thus, the M-H measurement indicates that introduction of nonmagnetic element into $TiO_2$ increases the magnetization of the samples, though not in a regular fashion because of different nature of substitution. Li and Mg doped compound lead to bulk magnetism, however $K^{+}/Ti^{4+}$ substitution lead to core-shell magnetism. The susceptibility as a function of temperature, measured in a field of 0.02T is shown in Figure 3(b). All samples display paramagnetic behavior. However, the plot of $\chi$T vs T data found to be almost linear, suggesting $\theta_P$ for all samples is almost zero. The effective paramagnetic moment, $\mu_{eff}$ was determined from the modified Curie-Weiss law, $\chi = \chi_0 + [(C/ (T- \theta_P)]$ and it was found to be 0.20, 0.26, 0.30 $\mu_B$ / ion for Li, Mg and K doped compounds respectively.

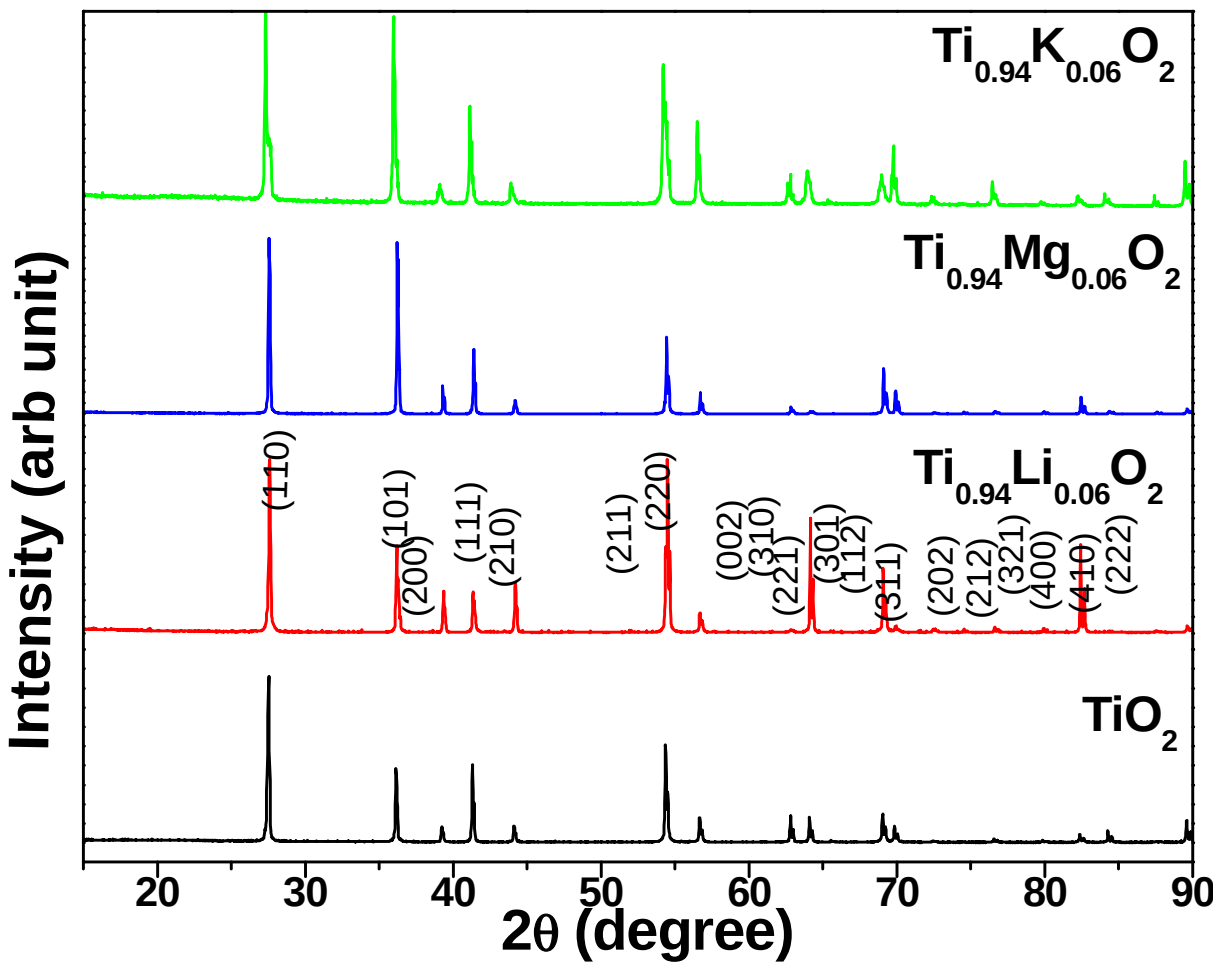

**FIGURE 1.** X-Ray Diffraction patterns of $Ti_{0.94}A_{0.06}O_2$ (A= Li, Mg, K) compounds.

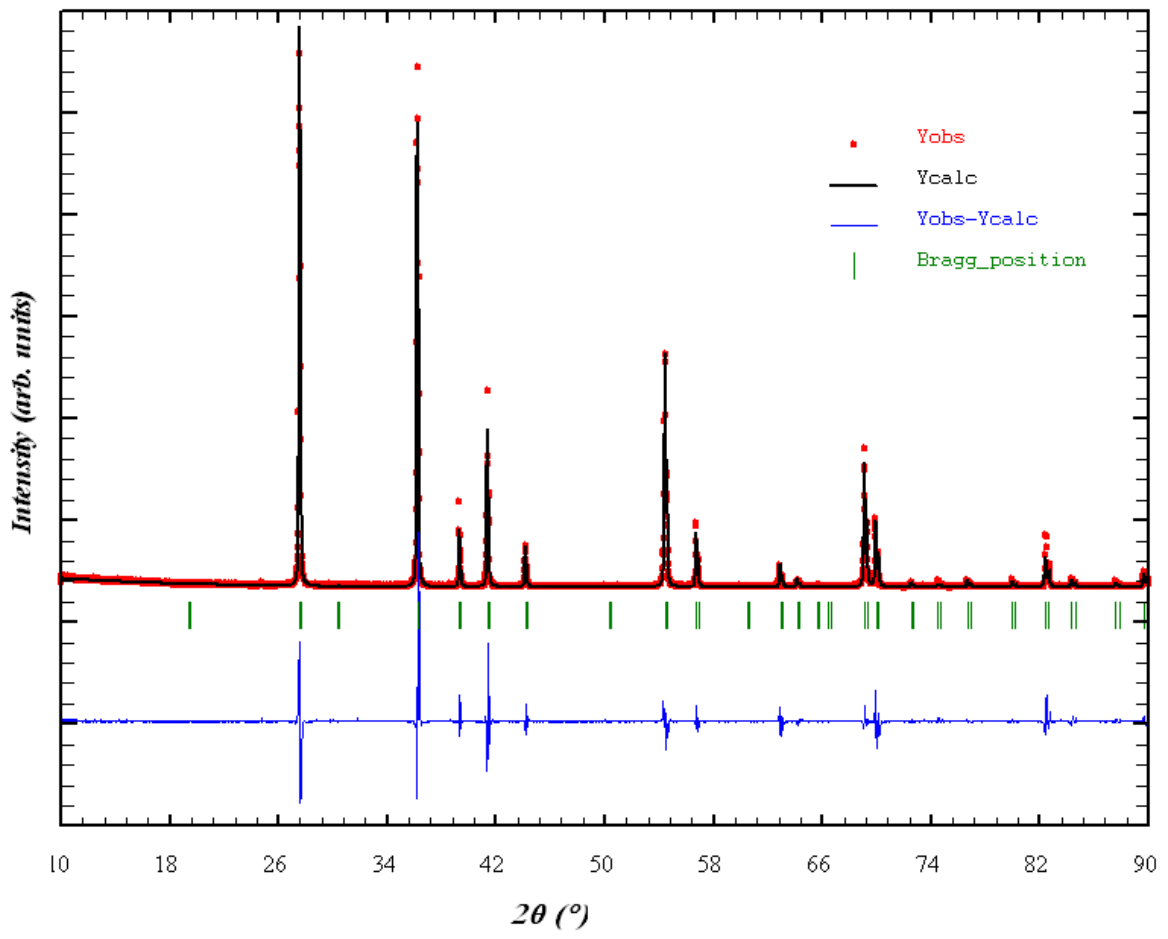

**FIGURE 2.** Refinement of X-Ray Diffraction patterns of $Ti_{0.94}Mg_{0.06}O_2$ compound.

Now, let us discuss the observed magnetism in these compounds in the light of existing first-principles calculations.

(i) *Bulk magnetism*: In the case of a direct cationic substitution, theoretical studies [5, 9] have demonstrated that three physical parameters are essential to explain induced $d^0$ magnetism: (i) the position of the induced impurity band which should be located near the top of the valence band, (ii) the density of carrier per defect, and (iii) the electron-electron correlations. The substitution of $Ti^{4+}$ by $Li^{+}$ and $Mg^{2+}$ in pure $TiO_2$ provides three/two holes in the present case respectively. In a recent first-principles calculation for Li doped $TiO_2$, it has been shown that Li substitution induces magnetism in $TiO_2$ [12]. The calculations indicate that the magnetic properties can be driven by the holes in O 2*p* orbitals which are induced by cation-site Li doping and delocalized distributed in the lattice. Indeed, the oxygen atoms surrounding the Li ion provide the dominant contribution to the total magnetic moment. *(ii) Core-Shell Magnetism in K doped $TiO_2$:* For K doped $TiO_2$ ab-initio study has predicted a large magnetic moment of $3\mu_B$ [13] comparable to that obtained in other samples of K and Na doped $ZrO_2$ [5]. In the ab initio studies one assumes a real substitution of $Ti^{4+}$ by $K^+$. The essential point is that thermodynamics is ignored in ab initio studies. Moreover, all previous theoretical calculations have focused mainly on bulk magnetism. Experimentally, we have seen that $K^+/Ti^{4+}$ substitution is not thermodynamically possible due to the large difference between charge or/and radius size. However, the introduction of K into $TiO_2$ induces some magnetism but the magnetic moment is far less than theoretically predicted and in particular, there is no long range ferromagnetic ordering.

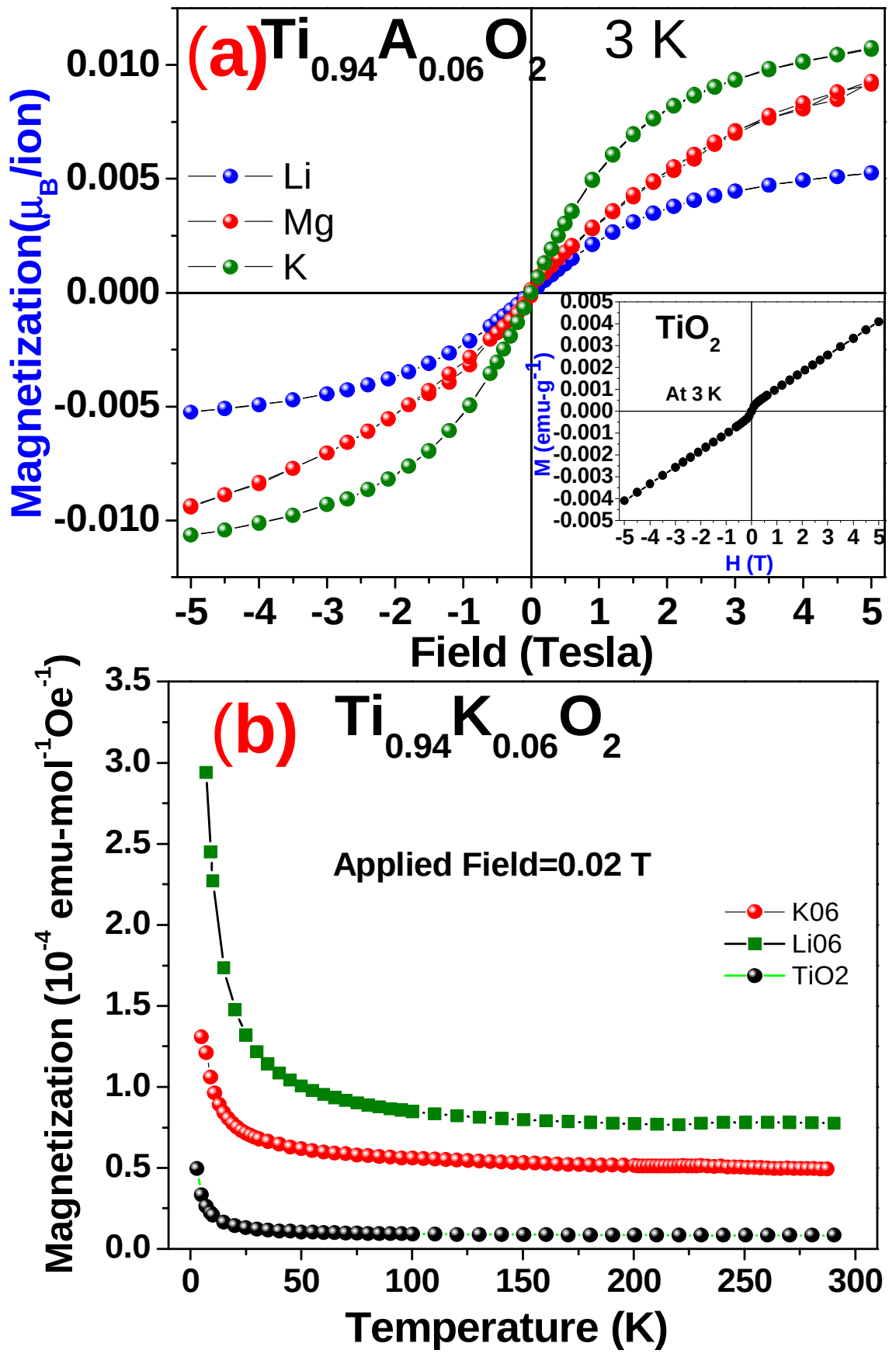

**FIGURE 3. (a)** M-H loops recorded at 3 K for $Ti_{0.94}A_{0.06}O_2$ (A=Li, Mg, K) compounds (b) Temperature variation of susceptibility.

# CONCLUSION

The $Ti_{0.94}A_{0.06}O_2$ (A=Li, Mg, K) compounds have been synthesized via solid state route method *at equilibrium*. The X-ray diffraction analyse of $Ti_{0.94}A_{0.06}O_2$ compounds provide the evidence that Li and Mg doped compound lead to Ti substitution. However K doped sample leads to core-shell structure. Weak but clear paramagnetic behaviour is observed for all sample and they exhibit different value of magnetism due to the difference in the nature of substitution.